# ASYMPTOTICS FOR GREEKS UNDER THE CONSTANT ELASTICITY OF VARIANCE MODEL

***Oleg L. Kritski*** [1]**, *Vladimir F. Zalmezh***

*Tomsk Polytechnic University, Russia*

**Abstract**

This paper is concerned with the asymptotics for Greeks of European-style options and the risk-neutral density function calculated under the constant elasticity of variance model. Formulae obtained help financial engineers to construct a perfect hedge with known behaviour and to price any options on financial assets.

## 1. Introduction

The computation of marginal behaviour of Greeks (or, the same, derivatives of option fair values) is important in financial risk management. Greeks represent the sensitivity of the price of derivative securities with respect to changes in the underlying asset prices $S_t$ or some major parameters like risk-free interest rate $r$, strike price $K$, volatility $\sigma$, expiration time $T$, time to maturity $\tau$. So, the information about behaviour of Greeks has practical and theoretical importance, especially if managers must hedge their securities for reducing risk, when future underlying price has no good assessment, or estimate the quality of the management of option strategy chosen.

Greeks and their asymptotics employ for many different ways such as for profit and loss attribution, and exotic contract design, and to calibrate parameters from market prices. For instance, it can be used for finding Black-Scholes implied variance behaviour [11, eq. (3.5), p. 28]:

$$\sigma^2(K,T)=\frac{1}{T}\int_0^T \frac{E(\sigma_t^2 S_t^2 \Gamma_C | F_0)}{E(S_t^2 \Gamma_C | F_0)}dt ,$$

where $\Gamma_C = \dfrac{\partial^2 C_t}{\partial S_t^2}$ is a Greek gamma for European-style call option with price $C_t = C_t(S_t)$, $\sigma_t$ is a local volatility, $(F_t)_{t\in[0,T]}$ is a filtration, generated by standard Brownian motion $W_t$, and $E$ is an expected value.

[1] Corresponding author. Tomsk Polytechnic University, 30 Lenin Ave., Tomsk, 634050, Russia.
Tel:+73822 606-335. Fax: +73822 705-691.
E-mail address: olegkol@tpu.ru

The marginal behaviour of Greeks is also useful for computing near-future price $C_t$ at known present price $C_{t-1}$ from Taylor series

$$dC_t \approx \Delta C_t = C_t - C_{t-1} = \Delta_C \Delta S_t + \frac{1}{2}\Gamma_C (\Delta S_t)^2 + \ldots,$$

where $\Delta_C = \frac{\partial C_t}{\partial S_t}$ is a Greek delta for call option. Furthermore, Black-Scholes equation [12] gives us a relationship between Greeks and call option price

$$r C_t = \Theta_C + \frac{\sigma^2 S_t^2}{2}\Gamma_C + r S_t \Delta_C,$$

where $\Theta_C = \frac{\partial C_t}{\partial t}$ is a Greek theta for call option.

Not only the Greeks' asymptotics, but limiting behaviour of some other derivatives of option fair prices are wholesome in financial mathematics. This is true, e.g., for the risk-neutral density function $\varphi(S_0, S_T, T)$ for final price $S_T$, that aggregates all appropriate information regarding preferences of derivative holders and underlying price dynamics. As it is known [4], $\varphi(S_0, S_T, T)$ is proportional to the second derivative of call option price $C_t = C_t(S_0, K, T)$ with respect to the exercise price $K$:

$$\varphi(S_0, S_T, T) = e^{rT} \left. \frac{\partial^2 C_t(S_0, K, T)}{\partial K^2} \right|_{K=S_T}.$$

In this work we calculate Greeks' asymptotics (section 2) and compute risk-neutral density function (section 3) for European options under the constant elasticity of variance (in brief, CEV) model. It is a kind of stochastic volatility model for the value of underlying asset proposed in [7]. It takes into account the negative correlation between stock returns and realized stock volatility (leverage effect); and the negative correlation between the strike price and the implied volatility [10] while the classical Black–Scholes model [12] doesn't. Moreover, there is some evidence [19], that risk-neutral European call prices can be non-monotonic functions of the time to maturity $\tau$ in which increasing time $t$ initially increases the call's price but after some point starts to depress the value of this call, and hence, its price can substantially differ from standard results of Black-Scholes.

Because of its importance, there are lots of financial applications of the CEV model in a wide variety of contexts: pricing American-style options with extension of Barone-Adesi and Whaley formula [2] or with Laplace–Carson transforms [20], or with inverse Laplace transform [13];

asymptotic expansions of option prices [16]; static hedge of American options [6] and American-style knock-in options on defaultable stocks [15]; binomial tree approach [9]; modeling nonlinear multivariate interest rate processes based on time-varying copulas and CEV approach [5] and many others.

## 2. Asymptotics for Greeks under the CEV model

Under the probability measure $P$ of probability space ($\Omega$, $F_t$, $P$) with a filtration $(F_t)_{t\in[0,T]}$, generated by standard Brownian motion $W=W_t$, the CEV process assumes that the asset price $\{S=S_t, 0\le t\le T\}$ is described by the following stochastic differential equation [7]:

$$dS = \mu S\, dt + \delta\, S^{\beta/2}\, dW\,,$$

where $S(0)=S_0$ is known, $\mu$ is an expected return rate, $\delta$ is a volatility and $0<\beta<2$ is a scale coefficient or elasticity of volatility, that makes the local volatility $\sigma_t = \delta S^{\beta/2-1}$ decline when the asset price grows.

Note that case $\beta=2$ corresponds to Black–Scholes model, case $\beta=0$ is the absolute diffusion process and case $\beta=1$ is the square-root diffusion model, both of them are described in [8].

Let from now on $\nu = (2-\beta)^{-1}, \tau = T-t$ and $m = \exp(r(2-\beta)\tau)$. The CEV call option price $C_t$ with exercise price $K$ and expiration time $T$ is expressed [17] as

$$C_t = S\,Q(2y; 2+2\nu, 2x) - Ke^{-r\tau}(1 - Q(2x; 2\nu, 2y)), \tag{1}$$

where

$$x = mkS^{2-\beta},\ y = kK^{2-\beta},\ k = \frac{2r}{\delta^2(2-\beta)(m-1)}, \tag{2}$$

$Q(2y;2\nu,2x)$ is a complementary non-central $\chi^2$-distribution function with $2\nu$ degrees of freedom and non-central parameter $2x$. It may be represented [17] as

$$Q(2y;2\nu,2x) = \frac{1}{2}\int_{2y}^{\infty} e^{-(2x+u)/2}\left(\frac{u}{2x}\right)^{(\nu-1)/2} I_{\nu-1}\left(\sqrt{2xu}\right)du = \sum_{n=1}^{\infty} g(n,x)G(n+\nu-1,y),$$

where $g(n,x) = \dfrac{e^{-x}x^{n-1}}{\Gamma(n)}$ is a density of complementary gamma distribution function, $\Gamma(n)$ is a gamma function, $G(n+\nu-1,y)$ is a complementary gamma distribution function, $G(n+\nu-1,y) = \int_y^{\infty} g(n+\nu-1,t)dt$, and $I_q(z) = \left(\dfrac{z}{2}\right)^q \sum_{j=0}^{\infty} \dfrac{(z^2/4)^j}{j!\Gamma(q+j+1)}$ is the modified Bessel function of the first kind of order $q$.

The CEV put option price $P_t$ is easy to find from 'call-put' parity [14] as follows

$$P_t = Ke^{-r\tau}Q(2x;2\nu,2y) - S\left(1 - Q(2y;2+2\nu,2x)\right). \quad (3)$$

Closed-form solutions for computing Greeks of European-style options under the CEV model are written in [14]:

$$\Delta_C = \frac{\partial C_t}{\partial S} = Q(2y;2+2\nu,2x) + 2x(2-\beta)p(2y;4+2\nu,2x) - \frac{2x(2-\beta)}{S}Ke^{-r\tau}p(2x;2\nu,2y), \quad (4a)$$

$$\Delta_P = \frac{\partial P_t}{\partial S} = \Delta_C - 1, \quad (4b)$$

$$\Gamma_C = \frac{\partial^2 C_t}{\partial S^2} = \frac{2x(2-\beta)^2}{S}\left(\frac{3-\beta}{2-\beta} - x\right)p(2y;4+2\nu,2x) + \frac{2x^2(2-\beta)^2}{S}p(2y;6+2\nu,2x) +$$

$$+ \frac{2x^2(2-\beta)^2}{S^2}Ke^{-r\tau}p(2x;2\nu,2y) - \frac{2xy(2-\beta)^2}{S^2}Ke^{-r\tau}p(2x;2+2\nu,2y) = \Gamma_P = \frac{\partial^2 P}{\partial S^2}, \quad (5)$$

$$\Theta_C = \frac{\partial C_t}{\partial t} = -\frac{\partial C_t}{\partial \tau} = -Kre^{-r\tau}Q(2y;2-2\nu,2x) + \frac{2rx(2-\beta)}{m-1}\left(Sp(2y;4+2\nu,2x) -\right.$$

$$\left. - Ke^{-r\tau}p(2x;2\nu,2y)\right), \quad (6a)$$

$$\Theta_P = \frac{\partial P_t}{\partial t} = Kre^{-r\tau}Q(2x;2\nu,2y) + \frac{2rx(2-\beta)}{m-1}\left(Sp(2y;4+2\nu,2x) - Ke^{-r\tau}p(2x;2\nu,2y)\right), \quad (6b)$$

$$Vega_C = \frac{\partial C_t}{\partial \sigma} = Vega_P = \frac{\partial P_t}{\partial \sigma} = \frac{-4x}{\sigma_0}\left(Sp(2y;4+2\nu,2x) - Ke^{-r\tau}p(2x;2\nu,2y)\right), \quad (7)$$

$$\rho_C = \frac{\partial C_t}{\partial r} = K\tau e^{-r\tau}\left(1 - Q(2x;2\nu,2y)\right) + 2x\left(\frac{1}{r} - \frac{(2-\beta)\tau}{m-1}\right)\left(Sp(2y;4+2\nu,2x) - Ke^{-r\tau}p(2x;2\nu,2y)\right), \quad (8a)$$

$$\rho_P = \frac{\partial C_t}{\partial r} = -K\tau e^{-r\tau}Q(2x;2\nu,2y) + 2x\left(\frac{1}{r} - \frac{(2-\beta)\tau}{m-1}\right)\left(Sp(2y;4+2\nu,2x) - Ke^{-r\tau}p(2x;2\nu,2y)\right), \quad (8b)$$

where $p(\omega;\nu,\lambda) = \frac{1}{2}e^{-(\lambda+\omega)/2}(\omega/\lambda)^{(\nu-2)/4}I_{(\nu-2)/2}\left(\sqrt{\lambda\omega}\right)$, ω>0, is a complementary non-central $\chi^2$–density function.

Let us examine the marginal behaviour of European-style call and put option prices, and their Greeks in eqs. (1), (3)–(8b), when $\tau \to 0$, $\sigma \to \infty$, $K \to \infty$, $r \to \infty$, $T \to \infty$.

**Case a.** $\tau \to 0$.

From eq. (2) it is obvious, that $\lim_{\tau\to 0} k = \lim_{\tau\to 0} x = \lim_{\tau\to 0} y = \infty$. In addition, $\lim_{\tau\to 0} m = 1$. For taking the limits needed we should discover the asymptotics for all the functions ingoing to expressions (1), (3)–(8b).

$$Q(2y;2+2\nu,2x) = \frac{1}{2}\int_{2y}^{\infty} e^{-(2x+u)/2}\left(\frac{u}{2x}\right)^{\nu/2} I_\nu\left(\sqrt{2xu}\right)du = (\text{sub } u=2xz^2) =$$

$$= 2x \int_{\sqrt{y/x}}^{\infty} e^{-x\left(1+z^2\right)} z^{\nu+1} I_\nu(2xz)\,dz\,, \tag{9}$$

$$Q(2x;2\nu,2y) = \frac{1}{2}\int_{2x}^{\infty} e^{-(2y+u)/2}\left(\frac{u}{2y}\right)^{(\nu-1)/2} I_{\nu-1}\left(\sqrt{2yu}\right)du = (\text{sub } u=2yz^2) =$$

$$= 2y \int_{\sqrt{x/y}}^{\infty} e^{-y\left(1+z^2\right)} z^{\nu} I_{\nu-1}(2yz)\,dz\,. \tag{10}$$

Further we use the modified Bessel function asymptotics when its argument $z \to \infty$, as it is given by [1, eq. (9.7.1)]:

$$I_\nu(z) \sim \frac{e^z}{\sqrt{2\pi z}}\,. \tag{11}$$

So, after substituting (11) into (9), (10) and completing the square, distribution functions are rewritten as

$$Q(2y;2+2\nu,2x) \sim \sqrt{\frac{x}{\pi}} \int_{\sqrt{y/x}}^{\infty} e^{-x(z-1)^2} z^{\nu+1/2} dz = (\text{sub } z = 1 + \frac{p}{\sqrt{2x}}) =$$

$$= \sqrt{\frac{1}{2\pi}} \int_{\sqrt{2x}\left(\sqrt{y/x}-1\right)}^{\infty} e^{-p^2/2}\left(1+\frac{p}{\sqrt{2x}}\right)^{\nu+1/2} dp = \sqrt{\frac{1}{2\pi}} \int_{\sqrt{2x}\left(\sqrt{y/x}-1\right)}^{\infty} e^{-p^2/2} dp + O\left(\frac{1}{\sqrt{2x}}\right); \tag{12}$$

$$Q(2x;2\nu,2y) = \sqrt{\frac{y}{\pi}} \int_{\sqrt{x/y}}^{\infty} e^{-y(z-1)^2} z^{\nu-1/2} dz = (\text{sub } z = 1 + \frac{p}{\sqrt{2y}}) =$$

$$= \sqrt{\frac{1}{2\pi}} \int_{\sqrt{2y}\left(\sqrt{x/y}-1\right)}^{\infty} e^{-p^2/2}\left(1+\frac{p}{\sqrt{2y}}\right)^{\nu-1/2} dp = \sqrt{\frac{1}{2\pi}} \int_{\sqrt{2y}\left(\sqrt{x/y}-1\right)}^{\infty} e^{-p^2/2} dp + O\left(\frac{1}{\sqrt{2y}}\right). \tag{13}$$

From eq. (13) we obtain

$$Q(2y;2-2\nu,2x) \sim 1 - \sqrt{\frac{1}{2\pi}} \int_{\sqrt{2y}\left(\sqrt{x/y}-1\right)}^{\infty} e^{-p^2/2} dp + O\left(\frac{1}{\sqrt{2y}}\right).$$

Using eq. (11) we can discover all the asymptotics of density distribution functions ingoing to (4a)–(8b) when $\tau \to 0$:

$$p(2x;2\nu;2y) \sim e^{-x} e^{-y} e^{2\sqrt{x}\sqrt{y}} C \frac{1}{\sqrt[4]{xy}} \sim \frac{C}{e^{k\left(\sqrt{mS^{2-\beta}}-\sqrt{E^{2-\beta}}\right)^2}\sqrt[4]{xy}}\,, \tag{14a}$$

$$p(2y;4+2\nu,2x) \sim e^{-x} e^{-y} e^{2\sqrt{x}\sqrt{y}} C_1 \frac{1}{\sqrt[4]{xy}} \sim \frac{C_1}{\sqrt[4]{xy}} e^{-\left(\sqrt{x}-\sqrt{y}\right)^2} \sim \frac{C_1}{e^{k\left(\sqrt{mS^{2-\beta}}-\sqrt{E^{2-\beta}}\right)^2}\sqrt[4]{xy}}\,, \tag{14b}$$

$$p(2y;4+2\nu,2x) \sim p(2y;6+2\nu,2x)\,, \tag{14c}$$

$$p(2x;2\nu;2y) \sim p(2x;2+2\nu;2y)\,, \tag{14d}$$

where $C$ and $C_1$ are constants, that independent from τ.

Taking the limits in (1), (3)–(8b) and using eq. (2) and asymptotics (12)–(14d) we get

$$\lim_{\tau\to 0} C_t = S\begin{cases}0, S<K\\1, S>K\end{cases} - K\begin{cases}0, S<K\\1, S>K\end{cases} = \begin{cases}0, S<K\\(S-K), S>K\end{cases};$$

$$\lim_{\tau\to 0} P_t = K\begin{cases}1, S<K\\0, S>K\end{cases} - S\begin{cases}1, S<K\\0, S>K\end{cases} = \begin{cases}(K-S), S<K\\0, S>K\end{cases};$$

$$\lim_{\tau\to 0}\Delta_C = \begin{cases}0, S<K\\1, S>K\end{cases} + 2(2-\beta)\lim_{\tau\to 0}\frac{C_1 x}{e^{k\left(\sqrt{mS^{2-\beta}}-\sqrt{E^{2-\beta}}\right)^2}\sqrt[4]{xy}} -$$

$$-\frac{2(2-\beta)K}{S}\lim_{\tau\to 0}\frac{Cx}{e^{k\left(\sqrt{mS^{2-\beta}}-\sqrt{E^{2-\beta}}\right)^2}\sqrt[4]{xy}} = \begin{cases}0, S<K\\1, S>K\end{cases};$$

$$\lim_{\tau\to 0}\Delta_P = \lim_{\tau\to 0}\Delta_C - 1 = \begin{cases}-1, S<K\\0, S>K\end{cases};$$

$$\lim_{\tau\to 0}\Gamma_C = \lim_{\tau\to 0}\Gamma_P = \lim_{\tau\to 0}\left[\frac{2x(2-\beta)^2}{S}\left(\frac{3-\beta}{2-\beta}-x\right)C_1 + \frac{2x^2(2-\beta)^2}{S}C_1 + \frac{2Kx^2(2-\beta)^2}{S^2}C - \right.$$

$$\left. -\frac{2Kxy(2-\beta)^2}{S^2}C\right]\frac{1}{e^{k\left(\sqrt{mS^{2-\beta}}-\sqrt{E^{2-\beta}}\right)^2}\sqrt[4]{xy}} = 0;$$

$$\lim_{\tau\to 0}\Theta_C = -rK\begin{cases}0, S<K\\1, S>K\end{cases} + \lim_{\tau\to 0}\left(\frac{2rx(2-\beta)(SC_1-Ke^{-r\tau}C)}{(m-1)e^{k\left(\sqrt{mS^{2-\beta}}-\sqrt{E^{2-\beta}}\right)^2}\sqrt[4]{xy}}\right) = -rK\begin{cases}0, S<K\\1, S>K\end{cases};$$

$$\lim_{\tau\to 0}\Theta_P = rK\begin{cases}1, S<K\\0, S>K\end{cases};$$

$$\lim_{\tau\to 0} Vega_C = \lim_{\tau\to 0} Vega_P = \frac{-4}{\sigma_0}\lim_{\tau\to 0}\left(x\frac{SC_1-Ke^{-r\tau}C}{e^{k\left(\sqrt{mS^{2-\beta}}-\sqrt{E^{2-\beta}}\right)^2}\sqrt[4]{xy}}\right) = 0.$$

Because of the finiteness of the limit $\lim_{\tau\to 0} Q(2x;2\nu;2y) = \begin{cases}1, S<K\\0, S>K\end{cases}$, we get

$$\lim_{\tau\to 0}\rho_C = 0 + 2\lim_{\tau\to 0}\left\{x\left(\frac{1}{r}-\frac{(2-\beta)\tau}{m-1}\right)\left(\frac{SC_1-Ke^{-r\tau}C}{e^{k\left(\sqrt{mS^{2-\beta}}-\sqrt{E^{2-\beta}}\right)^2}\sqrt[4]{xy}}\right)\right\} = 0;$$

$$\lim_{\tau\to 0}\rho_P = 0.$$

**Case b.** $\sigma\to\infty$.

From eq. (2) it is obvious, that $\lim_{\sigma\to\infty} k = \lim_{\sigma\to\infty} x = \lim_{\sigma\to\infty} y = 0$. From definition of CEV model it follows that $\sigma^2 = \delta^2 S^{\beta-2}$, $0<\beta<2$, i.e. $\delta\to\infty$ as $\sigma\to\infty$.

Like in the case $\tau \to 0$ we find the asymptotic behaviour of distribution functions ingoing to (1), (3), (6a) when $\sigma \to \infty$. We use eqs. (9), (10) and the modified Bessel function asymptotics when its argument $z \to 0$, as it is given by [1, eq. (9.6.7)]

$$I_\nu(2z) \sim \frac{z^\nu}{\Gamma(\nu+1)}. \tag{15}$$

Then

$$Q(2y;2+2\nu;2x) \sim 2\frac{x^{\nu+1}e^{-x}}{\Gamma(\nu+1)} \int_{\sqrt{y/x}}^{\infty} e^{-xz^2} z^{2\nu+1}\,dz = (\text{sub } z=\sqrt{t/x}) = \frac{e^{-x}}{\Gamma(\nu+1)}\int_y^\infty e^{-t}t^\nu dt; \tag{16}$$

$$Q(2x;2\nu,2y) \sim \frac{2y^\nu e^{-y}}{\Gamma(\nu)} \int_{\sqrt{x/y}}^{\infty} e^{-yz^2} z^{2\nu-1}dz = (\text{sub } z=\sqrt{t/y}) = \frac{e^{-y}}{\Gamma(\nu)}\int_x^\infty e^{-t}t^{\nu-1}dt. \tag{17}$$

Using (15) we discover the asymptotics of density distribution functions ingoing to (4a)–(8b) when $\sigma \to \infty$:

$$p(2y;4+2\nu,2x) \sim \frac{1}{2}\frac{e^{-x-y}}{\Gamma(\nu+2)} y^{\nu+1}, \tag{18a}$$

$$p(2x;2\nu,2y) \sim \frac{1}{2}\frac{e^{-x-y}}{\Gamma(\nu)} x^{\nu-1}, \tag{18b}$$

$$p(2y;6+2\nu,2x) \sim \frac{1}{2}\frac{e^{-x-y}}{\Gamma(\nu+3)} y^{\nu+2}, \tag{18c}$$

$$p(2x;2+2\nu,2y) \sim \frac{1}{2}\frac{e^{-x-y}}{\Gamma(\nu+1)} x^{\nu}. \tag{18d}$$

Taking the limits in (1), (3)–(8b) and using asymptotics (16)–(18d) we get

$$\lim_{\sigma\to\infty} C_t = S\frac{e^{-x}}{\Gamma(\nu+1)}\int_y^\infty e^{-t}t^\nu dt - Ke^{-r\tau}\left(1-\frac{e^{-y}}{\Gamma(\nu)}\int_x^\infty e^{-t}t^{\nu-1}dt\right) = S,$$

$$\lim_{\sigma\to\infty} P_t = \lim_{\sigma\to\infty}\left[Ke^{-r\tau}\frac{e^{-y}}{\Gamma(\nu)}\int_x^\infty e^{-t}t^{\nu-1}dt - S\left(1-\frac{e^{-x}}{\Gamma(\nu+1)}\int_y^\infty e^{-t}t^\nu dt\right)\right] = Ke^{-r\tau};$$

$$\lim_{\sigma\to\infty}\Delta_C = \lim_{\sigma\to\infty}\left(\frac{e^{-x}}{\Gamma(\nu+1)}\int_y^\infty e^{-t}t^\nu dt\right) + \lim_{\sigma\to\infty}\left(x(2-\beta)\frac{e^{-x-y}}{\Gamma(\nu+2)}y^{\nu+1}\right) -$$

$$-\lim_{\sigma\to\infty}\left(\frac{(2-\beta)}{S}Ke^{-r\tau}\frac{e^{-x-y}}{\Gamma(\nu)}x^\nu\right) = 1;$$

$$\lim_{\sigma\to\infty}\Delta_P = \lim_{\sigma\to\infty}\Delta_C - 1 = 0;$$

$$\lim_{\sigma\to\infty}\Gamma_C = \lim_{\sigma\to\infty}\Gamma_P = \lim_{\sigma\to\infty}\left[\frac{x(2-\beta)^2}{S}\left(\frac{3-\beta}{2-\beta}-x\right)\frac{e^{-x-y}}{\Gamma(\nu+2)}y^{\nu+1} + \right.$$

$$+\frac{x^2(2-\beta)^2}{S}\frac{e^{-x-y}}{\Gamma(\nu+3)}y^{\nu+2}+\frac{x^2(2-\beta)^2Ke^{-r\tau}}{S^2}\frac{e^{-x-y}}{\Gamma(\nu)}x^{\nu-1}-\frac{xy(2-\beta)^2Ke^{-r\tau}}{S^2}\frac{e^{-x-y}}{\Gamma(\nu+1)}x^{\nu}\Bigg]=0\,;$$

$$\lim_{\sigma\to\infty}\Theta_C=-Kre^{-r\tau}\lim_{\sigma\to\infty}\left(1-\frac{e^{-y}}{\Gamma(\nu)}\int_x^\infty e^{-t}t^{\nu-1}dt\right)+\frac{r(2-\beta)}{m-1}\lim_{\sigma\to\infty}\left(xS\frac{e^{-x-y}}{\Gamma(\nu+2)}y^{\nu+1}-Ke^{-r\tau}\frac{e^{-x-y}}{\Gamma(\nu)}x^{\nu}\right)=0\,;$$

$$\lim_{\sigma\to\infty}\Theta_P=Kre^{-r\tau}\lim_{\sigma\to\infty}\left(\frac{e^{-y}}{\Gamma(\nu)}\int_x^\infty e^{-t}t^{\nu-1}dt\right)+\frac{r(2-\beta)}{m-1}\lim_{\sigma\to\infty}\left(xS\frac{e^{-x-y}}{\Gamma(\nu+2)}y^{\nu+1}-Ke^{-r\tau}\frac{e^{-x-y}}{\Gamma(\nu)}x^{\nu}\right)=Kre^{-r\tau}\,;$$

$$\lim_{\sigma\to\infty}Vega_C=\lim_{\sigma\to\infty}Vega_P=\frac{-2}{\sigma_0}\lim_{\sigma\to\infty}\left(xS\frac{e^{-x-y}}{\Gamma(\nu+2)}y^{\nu+1}-Ke^{-r\tau}\frac{e^{-x-y}}{\Gamma(\nu)}x^{\nu}\right)=0\,;$$

$$\lim_{\sigma\to\infty}\rho_C=K\tau e^{-r\tau}\lim_{\sigma\to\infty}\left(1-\frac{e^{-y}}{\Gamma(\nu)}\int_x^\infty e^{-t}t^{\nu-1}dt\right)+\left(\frac{1}{r}-\frac{(2-\beta)\tau}{m-1}\right)\lim_{\sigma\to\infty}\left(xS\frac{e^{-x-y}}{\Gamma(\nu+2)}y^{\nu+1}-\right.$$

$$\left.-Ke^{-r\tau}\frac{e^{-x-y}}{\Gamma(\nu)}x^{\nu}\right)=0\,;$$

$$\lim_{\sigma\to\infty}\rho_P=-K\tau e^{-r\tau}\lim_{\sigma\to\infty}\left(\frac{e^{-y}}{\Gamma(\nu)}\int_x^\infty e^{-t}t^{\nu-1}dt\right)+\left(\frac{1}{r}-\frac{(2-\beta)\tau}{m-1}\right)\lim_{\sigma\to\infty}\left(xS\frac{e^{-x-y}}{\Gamma(\nu+2)}y^{\nu+1}-\right.$$

$$\left.-Ke^{-r\tau}\frac{e^{-x-y}}{\Gamma(\nu)}x^{\nu}\right)=-K\tau e^{-r\tau}\,.$$

**Case c.** $K\to\infty$.

From eq. (2) it is obvious, that $\lim\limits_{K\to\infty}k=k=\text{const},\ \lim\limits_{K\to\infty}x=x=\text{const},\ \lim\limits_{K\to\infty}y=\infty$. Therefore we use eqs. (12), (13) as distribution function asymptotics and eq. (11) as the modified Bessel function asymptotics. We have

$$p(2y;4+2\nu,2x)\sim\frac{1}{4\sqrt{\pi}}\frac{e^{-\left(\sqrt{x}-\sqrt{y}\right)^2}}{\sqrt[4]{xy}}\left(\frac{y}{x}\right)^{(\nu+1)/2}, \tag{19a}$$

$$p(2y;6+2\nu,2x)\sim\frac{1}{4\sqrt{\pi}}\frac{e^{-\left(\sqrt{x}-\sqrt{y}\right)^2}}{\sqrt[4]{xy}}\left(\frac{y}{x}\right)^{(\nu+2)/2}, \tag{19b}$$

$$p(2x;2\nu,2y)\sim\frac{1}{4\sqrt{\pi}}\frac{e^{-\left(\sqrt{x}-\sqrt{y}\right)^2}}{\sqrt[4]{xy}}\left(\frac{x}{y}\right)^{(\nu-1)/2}, \tag{19c}$$

$$p(2x;2+2\nu,2y)\sim\frac{1}{4\sqrt{\pi}}\frac{e^{-\left(\sqrt{x}-\sqrt{y}\right)^2}}{\sqrt[4]{xy}}\left(\frac{x}{y}\right)^{\nu/2}. \tag{19d}$$

Taking the limits in (1), (3)–(8b) and using asymptotics (12), (13), (19a)–(19d) we get

$$\lim_{K\to\infty}C_t=S\lim_{K\to\infty}\left(\sqrt{\frac{1}{2\pi}}\int_{\sqrt{2y}-\sqrt{2x}}^\infty e^{-p^2/2}dp\right)-e^{-r\tau}\lim_{K\to\infty}K\left(1-\sqrt{\frac{1}{2\pi}}\int_{\sqrt{2x}-\sqrt{2y}}^\infty e^{-p^2/2}dp\right)=$$

$$= -e^{-r\tau}\lim_{K\to\infty} K\left(1 - 1 + \sqrt{\frac{1}{2\pi}}\int_{-\infty}^{\sqrt{2x}-\sqrt{2y}} e^{-p^2/2}dp\right) = e^{-r\tau}\lim_{K\to\infty} K\,\frac{e^{-\left(\sqrt{2x}-\sqrt{2y}\right)^2/2}}{\sqrt{2\pi}\left(\sqrt{2x}-\sqrt{2y}\right)} = 0,$$

$$\text{as } \sqrt{\frac{1}{2\pi}}\int_{-\infty}^{\alpha} e^{-p^2/2}dp \sim -\frac{e^{-\alpha^2/2}}{\sqrt{2\pi}\alpha} \text{ when } \alpha \to -\infty;$$

$$\lim_{K\to\infty} P_t = e^{-r\tau}\lim_{K\to\infty} K\left(\sqrt{\frac{1}{2\pi}}\int_{\sqrt{2x}-\sqrt{2y}}^{\infty} e^{-p^2/2}dp\right) - S\lim_{K\to\infty}\left(1 - \sqrt{\frac{1}{2\pi}}\int_{\sqrt{2y}-\sqrt{2x}}^{\infty} e^{-p^2/2}dp\right) = \infty;$$

$$\lim_{K\to\infty}\Delta_C = \lim_{K\to\infty}\left(\sqrt{\frac{1}{2\pi}}\int_{\sqrt{2y}-\sqrt{2x}}^{\infty} e^{-p^2/2}dp\right) + \frac{x(2-\beta)}{2\sqrt{\pi}}\lim_{K\to\infty}\frac{e^{-\left(\sqrt{x}-\sqrt{y}\right)^2}}{\sqrt[4]{xy}}\left(\frac{y}{x}\right)^{(\nu+1)/2} -$$

$$-\frac{x(2-\beta)e^{-r\tau}}{2\sqrt{\pi}S}\lim_{K\to\infty} K\,\frac{e^{-\left(\sqrt{x}-\sqrt{y}\right)^2}}{\sqrt[4]{xy}}\left(\frac{x}{y}\right)^{(\nu-1)/2} = 0;$$

$$\lim_{K\to\infty}\Delta_P = \lim_{K\to\infty}\Delta_C - 1 = -1;$$

$$\lim_{K\to\infty}\Gamma_C = \lim_{K\to\infty}\Gamma_P = \frac{2x(2-\beta)^2}{S}\left(\frac{3-\beta}{2-\beta} - x\right)\lim_{K\to\infty} p(2y;4+2\nu,2x) + \frac{2x^2(2-\beta)^2}{S}\lim_{K\to\infty} p(2y;6+2\nu,2x) +$$

$$+\frac{x^2(2-\beta)^2 e^{-r\tau}}{2\sqrt{\pi}S^2}\lim_{K\to\infty} K\,\frac{e^{-\left(\sqrt{x}-\sqrt{y}\right)^2}}{\sqrt[4]{xy}}\left(\frac{x}{y}\right)^{(\nu-1)/2} - \frac{xe^{-r\tau}(2-\beta)^2}{2\sqrt{\pi}S^2}\lim_{K\to\infty} yK\,\frac{e^{-\left(\sqrt{x}-\sqrt{y}\right)^2}}{\sqrt[4]{xy}}\left(\frac{x}{y}\right)^{\nu/2} = 0;$$

$$\lim_{K\to\infty}\Theta_C = -re^{-r\tau}\lim_{K\to\infty} K\left(1 - \sqrt{\frac{1}{2\pi}}\int_{\sqrt{2x}-\sqrt{2y}}^{\infty} e^{-p^2/2}dp\right) +$$

$$+\frac{rx(2-\beta)}{2\sqrt{\pi}(m-1)}\lim_{K\to\infty}\frac{e^{-\left(\sqrt{x}-\sqrt{y}\right)^2}}{\sqrt[4]{xy}}\left(S\left(\frac{y}{x}\right)^{(\nu+1)/2} - Ke^{-r\tau}\left(\frac{x}{y}\right)^{(\nu-1)/2}\right) =$$

$$= -re^{-r\tau}\lim_{K\to\infty} K\left(\frac{e^{-\left(\sqrt{2x}-\sqrt{2y}\right)^2/2}}{\sqrt{2\pi}\left(\sqrt{2x}-\sqrt{2y}\right)}\right) = 0,$$

$$\text{as } \sqrt{\frac{1}{2\pi}}\int_{-\infty}^{\alpha} e^{-p^2/2}dp \sim -\frac{e^{-\alpha^2/2}}{\sqrt{2\pi}\alpha} \text{ when } \alpha \to -\infty;$$

$$\lim_{K\to\infty}\Theta_P = re^{-r\tau}\lim_{K\to\infty} K\left(\sqrt{\frac{1}{2\pi}}\int_{\sqrt{2x}-\sqrt{2y}}^{\infty} e^{-p^2/2}dp\right) = \infty;$$

$$\lim_{K\to\infty} Vega_C = \lim_{K\to\infty} Vega_P = \frac{-x}{\sqrt{\pi}\sigma_0}\lim_{K\to\infty}\frac{e^{-\left(\sqrt{x}-\sqrt{y}\right)^2}}{\sqrt[4]{xy}}\left(S\left(\frac{y}{x}\right)^{(\nu+1)/2} - Ke^{-r\tau}\left(\frac{x}{y}\right)^{(\nu-1)/2}\right) = 0;$$

$$\lim_{K\to\infty}\rho_C = \tau e^{-r\tau}\lim_{K\to\infty} K\left(1 - \sqrt{\frac{1}{2\pi}}\int_{\sqrt{2x}-\sqrt{2y}}^{\infty} e^{-p^2/2}dp\right) +$$

$$+\frac{x}{2\sqrt{\pi}}\left(\frac{1}{r}-\frac{(2-\beta)\tau}{m-1}\right)\lim_{K\to\infty}\frac{e^{-\left(\sqrt{x}-\sqrt{y}\right)^2}}{\sqrt[4]{xy}}\left(S\left(\frac{y}{x}\right)^{(\nu+1)/2}-Ke^{-r\tau}\left(\frac{x}{y}\right)^{(\nu-1)/2}\right)=$$

$$=\frac{\tau e^{-r\tau}}{\sqrt{2\pi}}\lim_{K\to\infty}\left(K\frac{e^{-\left(\sqrt{2x}-\sqrt{2y}\right)^2/2}}{\left(\sqrt{2x}-\sqrt{2y}\right)}\right)=0,$$

$$\text{as }\sqrt{\frac{1}{2\pi}}\int_{-\infty}^{\alpha}e^{-p^2/2}dp\sim-\frac{e^{-\alpha^2/2}}{\sqrt{2\pi}\alpha}\text{ when }\alpha\to-\infty;$$

$$\lim_{K\to\infty}\rho_P=-\tau e^{-r\tau}\lim_{K\to\infty}K\left(\sqrt{\frac{1}{2\pi}}\int_{\sqrt{2x}-\sqrt{2y}}^{\infty}e^{-p^2/2}dp\right)=-\infty;$$

**Case d.** $r\to\infty$.

From eq. (2) it is obvious, that $\lim_{r\to\infty}k=0,\lim_{r\to\infty}x=\infty,\lim_{r\to\infty}y=0,\lim_{r\to\infty}xy=0$. In addition, $\lim_{r\to\infty}m=\infty$. Therefore we have the same asymptotics as in eqs. (12), (17), (18a)–(18d). So, we get

$$\lim_{r\to\infty}C_t=S\lim_{r\to\infty}\left(\sqrt{\frac{1}{2\pi}}\int_{\sqrt{2y}-\sqrt{2x}}^{\infty}e^{-p^2/2}dp\right)-K\lim_{r\to\infty}e^{-r\tau}\left(1-\frac{e^{-y}}{\Gamma(\nu)}\int_x^{\infty}e^{-t}t^{\nu-1}dt\right)=S;$$

$$\lim_{r\to\infty}P_t=K\lim_{r\to\infty}\left(e^{-r\tau}\frac{e^{-y}}{\Gamma(\nu)}\int_x^{\infty}e^{-t}t^{\nu-1}dt\right)-S\lim_{r\to\infty}\left(1-\sqrt{\frac{1}{2\pi}}\int_{\sqrt{2y}-\sqrt{2x}}^{\infty}e^{-p^2/2}dp\right)=0;$$

$$\lim_{r\to\infty}\Delta_C=\lim_{r\to\infty}\left(\sqrt{\frac{1}{2\pi}}\int_{\sqrt{2y}-\sqrt{2x}}^{\infty}e^{-p^2/2}dp\right)+(2-\beta)\lim_{r\to\infty}\left(\frac{e^{-x-y}x}{\Gamma(\nu+2)}y^{\nu+1}\right)-\frac{K(2-\beta)}{S}\lim_{r\to\infty}\left(e^{-r\tau}\frac{e^{-x-y}}{\Gamma(\nu)}x^{\nu}\right)=1;$$

$$\lim_{r\to\infty}\Delta_P=\lim_{r\to\infty}\Delta_C-1=0;$$

$$\lim_{r\to\infty}\Gamma_C=\lim_{r\to\infty}\Gamma_P=\lim_{r\to\infty}\left(\frac{x(2-\beta)^2}{S}\left(\frac{3-\beta}{2-\beta}-x\right)\frac{e^{-x-y}}{\Gamma(\nu+2)}y^{\nu+1}\right)+\lim_{r\to\infty}\left(\frac{x^2(2-\beta)^2}{S}\frac{e^{-x-y}}{\Gamma(\nu+3)}y^{\nu+2}\right)+$$

$$+\lim_{r\to\infty}\left(\frac{(2-\beta)^2}{S^2}Ke^{-r\tau}\frac{e^{-x-y}}{\Gamma(\nu)}x^{\nu+1}\right)-\lim_{r\to\infty}\left(\frac{y(2-\beta)^2}{S^2}Ke^{-r\tau}\frac{e^{-x-y}}{\Gamma(\nu+1)}x^{\nu+1}\right)=0;$$

$$\lim_{r\to\infty}\Theta_C=-K\lim_{r\to\infty}re^{-r\tau}\left(1-\frac{e^{-y}}{\Gamma(\nu)}\int_x^{\infty}e^{-t}t^{\nu-1}dt\right)+\lim_{r\to\infty}\frac{rx(2-\beta)}{m-1}\left(S\frac{e^{-x-y}}{\Gamma(\nu+2)}y^{\nu+1}-Ke^{-r\tau}\frac{e^{-x-y}}{\Gamma(\nu)}x^{\nu-1}\right)=0;$$

$$\lim_{r\to\infty}\Theta_P=K\lim_{r\to\infty}re^{-r\tau}\left(\frac{e^{-y}}{\Gamma(\nu)}\int_x^{\infty}e^{-t}t^{\nu-1}dt\right)+\lim_{r\to\infty}\frac{rx(2-\beta)}{m-1}\left(S\frac{e^{-x-y}}{\Gamma(\nu+2)}y^{\nu+1}-Ke^{-r\tau}\frac{e^{-x-y}}{\Gamma(\nu)}x^{\nu-1}\right)=0;$$

$$\lim_{r\to\infty}Vega_C=\lim_{r\to\infty}Vega_P=-2\lim_{r\to\infty}\frac{x}{\sigma_0}\left(S\frac{e^{-x-y}}{\Gamma(\nu+2)}y^{\nu+1}-Ke^{-r\tau}\frac{e^{-x-y}}{\Gamma(\nu)}x^{\nu-1}\right)=0;$$

$$\lim_{r\to\infty}\rho_C = K\tau\lim_{r\to\infty} e^{-r\tau}\left(1-\frac{e^{-y}}{\Gamma(\nu)}\int_x^{\infty} e^{-t}t^{\nu-1}dt\right)+\lim_{r\to\infty} x\left(\frac{1}{r}-\frac{\tau(2-\beta)}{m-1}\right)\left(S\frac{e^{-x-y}y^{\nu+1}}{\Gamma(\nu+2)}-Ke^{-r\tau}\frac{e^{-x-y}x^{\nu-1}}{\Gamma(\nu)}\right)=0;$$

$$\lim_{r\to\infty}\rho_P = -K\tau\lim_{r\to\infty} e^{-r\tau}\left(\frac{e^{-y}}{\Gamma(\nu)}\int_x^{\infty} e^{-t}t^{\nu-1}dt\right)+\lim_{r\to\infty} x\left(\frac{1}{r}-\frac{\tau(2-\beta)}{m-1}\right)\left(S\frac{e^{-x-y}y^{\nu+1}}{\Gamma(\nu+2)}-Ke^{-r\tau}\frac{e^{-x-y}x^{\nu-1}}{\Gamma(\nu)}\right)=0.$$

**Case e.** $T\to\infty$.

From eq. (2) it is obvious, that $\lim_{T\to\infty} k=0, \lim_{T\to\infty} x=C, \lim_{T\to\infty} y=0$ and $\lim_{T\to\infty} xy=0$, where $C$ is some constant, that independent from *T*. In addition, $\lim_{T\to\infty} m=\infty$. Taking the limits in eqs. (9), (10) we get

$$Q(2y;2+2\nu,2x)\to 2C\int_0^{\infty} e^{-C(1+z^2)}z^{\nu+1}I_\nu(2Cz)dz = (\text{sub } z=\sqrt{t}\,) = Ce^{-C}\int_0^{\infty} e^{-Ct}t^{\nu/2}I_\nu\left(2C\sqrt{t}\right)dt; \quad (20)$$

$$Q(2x;2\nu,2y)=2y\int_{\sqrt{x/y}}^{\infty} e^{-y(1+z^2)}z^{\nu}I_{\nu-1}(2yz)dz \sim \frac{2y^{\nu}e^{-y}}{\Gamma(\nu)}\int_{\sqrt{x/y}}^{\infty} e^{-yz^2}z^{2\nu-1}dz = (\text{sub } z=p(2y)^{-1/2}) =$$

$$=\frac{e^{-y}}{2^{\nu-1}\Gamma(\nu)}\int_{\sqrt{2x}}^{\infty} e^{-p^2/2}p^{2\nu-1}dp \to \frac{1}{2^{\nu-1}\Gamma(\nu)}\int_{\sqrt{2C}}^{\infty} e^{-p^2/2}p^{2\nu-1}dp = \text{const}. \quad (21)$$

Note that integral in eq. (20) is a Laplace transform for the function $f(t)=t^{\nu/2}I_\nu\left(2C\sqrt{t}\right)$. It equals to $e^C/C$ as it is given by [3, eq. (18), p. 197]. On the whole $Q(2y;2+2\nu,2x)\to 1$.

For density distribution functions we have the same asymptotics as in eqs. (18a)–(18d), so, taking the limit $T\to\infty$ in (1), (3)–(8b) and using eqs. (20), (21) additionally, we have

$$\lim_{T\to\infty} C_t = S\lim_{T\to\infty} Q(2y;2+2\nu,2x) - K\lim_{T\to\infty} e^{-r\tau}\left(1-Q(2x;2\nu,2y)\right)=S;$$

$$\lim_{T\to\infty} P_t = K\lim_{T\to\infty} e^{-r\tau}Q(2x;2\nu,2y) - S\lim_{T\to\infty}\left(1-Q(2y;2+2\nu,2x)\right)=0;$$

$$\lim_{T\to\infty}\Delta_C = \lim_{T\to\infty} Q(2y;2+2\nu,2x)+(2-\beta)\lim_{T\to\infty}\frac{xe^{-x-y}y^{\nu+1}}{\Gamma(\nu+2)} -$$

$$-\frac{K(2-\beta)}{S}\lim_{T\to\infty} e^{-r\tau}\frac{e^{-x-y}}{\Gamma(\nu)}x^{\nu}=1;$$

$$\lim_{T\to\infty}\Delta_P = \lim_{T\to\infty}\Delta_C - 1 = 0;$$

$$\lim_{T\to\infty}\Gamma_C = \lim_{T\to\infty}\Gamma_P = \lim_{T\to\infty}\frac{x(2-\beta)^2}{S}\left(\frac{3-\beta}{2-\beta}-x\right)\frac{e^{-x-y}y^{\nu+1}}{\Gamma(\nu+2)} + \lim_{T\to\infty}\frac{x^2(2-\beta)^2}{S}\frac{e^{-x-y}y^{\nu+2}}{\Gamma(\nu+3)} +$$

$$+\lim_{T\to\infty}\frac{(2-\beta)^2}{S^2}Ke^{-r\tau}\frac{e^{-x-y}}{\Gamma(\nu)}x^{\nu+1} - \lim_{T\to\infty}\frac{y(2-\beta)^2}{S^2}Ke^{-r\tau}\frac{e^{-x-y}}{\Gamma(\nu+1)}x^{\nu+1}=0;$$

$$\lim_{T\to\infty}\Theta_C = -Kr\lim_{T\to\infty} e^{-r\tau}\left(1-Q(2x;2\nu,2y)\right)+r(2-\beta)\lim_{T\to\infty}\left(\frac{xSe^{-x-y}y^{\nu+1}}{\Gamma(\nu+2)(m-1)}-\frac{x^{\nu}Ke^{-x-y}e^{-r\tau}}{(m-1)\Gamma(\nu)}\right)=0\,;$$

$$\lim_{T\to\infty}\Theta_P = Kr\lim_{T\to\infty} e^{-r\tau}Q(2x;2\nu,2y)+r(2-\beta)\lim_{T\to\infty}\left(\frac{xSe^{-x-y}y^{\nu+1}}{\Gamma(\nu+2)(m-1)}-\frac{x^{\nu}Ke^{-x-y}e^{-r\tau}}{(m-1)\Gamma(\nu)}\right)=0\,;$$

$$\lim_{T\to\infty} Vega_C = \lim_{T\to\infty} Vega_P = -2\lim_{T\to\infty}\frac{x}{\sigma_0}\left(S\frac{e^{-x-y}}{\Gamma(\nu+2)}y^{\nu+1}-Ke^{-r\tau}\frac{e^{-x-y}}{\Gamma(\nu)}x^{\nu-1}\right)=0\,;$$

$$\lim_{T\to\infty}\rho_C = K\lim_{T\to\infty}\tau e^{-r\tau}\left(1-Q(2x;2\nu,2y)\right)+2\lim_{T\to\infty} x\left(\frac{1}{r}-\frac{(2-\beta)\tau}{m-1}\right)\left(Sp(2y;4+2\nu,2x)-Ke^{-r\tau}p(2x;2\nu,2y)\right)=$$

$$=\lim_{T\to\infty} x\left(\frac{1}{r}-\frac{(2-\beta)\tau}{m-1}\right)\left(S\frac{e^{-x-y}}{\Gamma(\nu+2)}y^{\nu+1}-Ke^{-r\tau}\frac{e^{-x-y}}{\Gamma(\nu)}x^{\nu-1}\right)=0\,;$$

$$\lim_{T\to\infty}\rho_P = -K\lim_{T\to\infty}\tau e^{-r\tau}Q(2x;2\nu,2y)=0\,.$$

## 3. Risk-neutral density function under the CEV model

Knowledge about the dynamics of the risk-neutral density is necessary for the pricing of any options on financial assets [4, eq. (3)], even exotic and complex [18].

**Theorem** 1: *under the CEV model there exists a risk-neutral density function* $\varphi(S_t,K,T)$ *for European-style options with the final price* $S_T=K$

$$\varphi(S_t,S_T,T)=e^{r\tau}\left.\frac{\partial^2 C_t(S_t,K,T)}{\partial K^2}\right|_{K=S_T},$$

where
$$\frac{\partial^2 C_t(S_t,K,T)}{\partial K^2}=\frac{2S(2-\beta)^2}{K^2}y(y-1+\nu)p(2y;2+2\nu,2x)-\frac{2S(2-\beta)}{K}y\,p(2y;2\nu,2x)+$$
$$+e^{-r\tau}\frac{2(2-\beta)^2}{K}y(1+\nu-y)p(2x;2+2\nu,2y)+e^{-r\tau}\frac{2(2-\beta)^2}{K}y^2\,p(2x;4+2\nu,2y).$$

**Proof**:

To compute $\varphi(S_t,S_T,T)$ we need use the following auxiliary relations as it is given by [14, eqs. (A2a), (A2b), (A11a), (A11b)]:

$$\frac{\partial Q(\omega;\nu,\lambda)}{\partial\omega}=-p(\omega;\nu,\lambda),\quad \frac{\partial Q(\omega;\nu,\lambda)}{\partial\lambda}=p(\omega;\nu+2,\lambda), \qquad (22)$$

$$\frac{\partial p(\omega;\nu,\lambda)}{\partial\omega}=\frac{1}{2}\left(-p(\omega;\nu,\lambda)+p(\omega;\nu-2,\lambda)\right),\quad \frac{\partial p(\omega;\nu,\lambda)}{\partial\lambda}=\frac{1}{2}\left(-p(\omega;\nu,\lambda)+p(\omega;\nu+2,\lambda)\right), \qquad (23)$$

Using eq. (22), we are able to compute the following partial derivatives:

$$\frac{\partial Q(2y;2+2\nu,2x)}{\partial K}=\frac{\partial(2y)}{\partial K}p(2y;2+2\nu,2x)=-\frac{2y(2-\beta)}{K}p(2y;2+2\nu,2x),$$

$$\frac{\partial Q(2x;2\nu,2y)}{\partial K}=\frac{\partial (2y)}{\partial K}p(2x;2+2\nu,2y)=\frac{2y(2-\beta)}{K}p(2x;2+2\nu,2y),$$

because $\frac{\partial (2x)}{\partial K}=0$, $\frac{\partial (2y)}{\partial K}=\frac{2y(2-\beta)}{K}$.

So, differentiating eq. (1), we have

$$\frac{\partial C_t}{\partial K}=-\frac{2S}{K}y(2-\beta)p(2y;2+2\nu,2x)-e^{-r\tau}(1-Q(2x;2\nu,2y))+2e^{-r\tau}y(2-\beta)p(2x;2+2\nu,2y).$$

Since $\frac{\partial}{\partial K}\left(\frac{y}{K}\right)=(1-\beta)\frac{y}{K^2}$, we use eq. (23) for computing the second partial derivative as follows:

$$\frac{\partial^2 C_t}{\partial K^2}=-\frac{2S}{K^2}y(1-\beta)(2-\beta)p(2y;2+2\nu,2x)-\frac{2S}{K^2}y^2(2-\beta)^2(-p(2y;2+2\nu,2x)+p(2y;2\nu,2x))+$$

$$+e^{-r\tau}\frac{2y(2-\beta)}{K}p(2x;2+2\nu,2y)+e^{-r\tau}\frac{2y(2-\beta)^2}{K}p(2x;2+2\nu,2y)+$$

$$+2e^{-r\tau}\frac{y^2(2-\beta)^2}{K}(-p(2x;2+2\nu,2y)+p(2x;4+2\nu,2y)).$$

Finally, reducing similar terms and simplifying we get

$$\frac{\partial^2 C_t}{\partial K^2}=\frac{2S(2-\beta)^2}{K^2}y(y-1+\nu)p(2y;2+2\nu,2x)-\frac{2S(2-\beta)}{K}y\,p(2y;2\nu,2x)+$$

$$+e^{-r\tau}\frac{2(2-\beta)^2}{K}y(1+\nu-y)p(2x;2+2\nu,2y)+e^{-r\tau}\frac{2(2-\beta)^2}{K}y^2p(2x;4+2\nu,2y).$$

This completes the proof.

**Acknowledgment**

The work is carried out at Tomsk Polytechnic University within the framework of Tomsk Polytechnic University Competitiveness Enhancement Program grant.